\documentclass[a4,12pt,titlepage]{article}
\makeatletter
\usepackage{amssymb}
\begin{document}

\begin{center}
{\bf\LARGE Space guaranteeing a primitive chaotic behavior}\\~\\~\\

{\bf\large Yoshihito Ogasawara}
\renewcommand{\thefootnote}{*}\footnote{
Corresponding author.\\
{\it E-mail address:} ogasawara@aoni.waseda.jp (Yoshihito Ogasawara).
}, {\bf\large Shin'ichi Oishi}
\end{center}

\begin{center}
Faculty of Science and Engineering, Waseda University, Ohkubo, Shinjuku-ku, Tokyo 169-8555, Japan
\end{center}

{\noindent\bf Abstract.}~~This study describes such a situation that a Cantor set emerges as a result of the exploration of sufficient conditions for the property which is generalized from fundamental chaotic maps, and the Cantor set even guarantees infinitely many varieties of the behavior with the property, as well as a typical continuum.\\

\section{Introduction}
\label{Introduction}

We are surrounded by chaotic behaviors and complex forms called fractals, and it is known that a fractal emerges in a chaotic behavior \cite{ENL,Smale,BBM,ASY,YU}. The purpose of this study is to reveal the relation of a fractal and a chaos from a new angle.  

Fundamental maps in the chaos theory, such as the tent map $\varphi:[0,1]\to[0,1],~x\mapsto\min\{2x,2(1-x)\}$, the baker map $\varphi:[0,1]\to[0,1],~x\mapsto 2x~(0\le x\le 1/2),~2x-1~(1/2\le x\le 1)$,  and the logistic function $\varphi:[0,1]\to[0,1],~x\mapsto 4x(1-x)$, have an intriguing property%
\begin{itemize}
\item[] ``For any infinite sequence $\omega_0,~\omega_1,~\omega_2,\ldots$, there exists an initial point $x_0\in\omega_0$ such that $\varphi(x_0)\in\omega_1,~\varphi(\varphi(x_0))\in\omega_2,\ldots$, where each $\omega_i$ is $[0,1/2]$ or $[1/2,1]$."
\end{itemize}
This property can play an important role in the guarantee of nonperiodicity, sensitive dependence on initial conditions, and so on for the maps \cite{ASY}. Then, it was pointed out \cite{O2010} that this property implies general problems about determinism, causality, free will, and so on, which are vital problems in sciences (e.g. Ref. \cite{EPR,JRS,CK,PSL,JM,MJ,RP}).

However, this property is too restricted in such a sense that the number of the subsets $[0,1/2]$ and $[1/2,1]$ which imply events or selections \cite{O2010} is only two and they are required for intersecting one point $1/2$. Then, the following property (P) generalized from this property was proposed as a primitive chaotic behavior, and it has been investigated \cite{O2010,CSF2006,CSF2008,O2011}. 
\begin{itemize}
\item[(P)] For any infinite sequence $\omega_0,~\omega_1,~\omega_2,\ldots$, there exists an initial point $x_0\in\omega_0$ such that $f_{\omega_0}(x_0)\in\omega_1,~f_{\omega_1}(f_{\omega_0}(x_0))\in\omega_2,\ldots$. Here, each $\omega_i$ is an element of a family $\{X_\lambda,~\lambda\in\Lambda\}$ of nonempty subsets of a set $X$, and each $f_{X_\lambda}$ is the map from $X_\lambda$ to $X$.
\end{itemize}
In a previous study \cite{O2010}, such a situation that a nondegenerate Peano continuum emerges by the exploration of sufficient conditions for the property (P) was described, and the following theorem was exhibited as a result of it, where a nondegenerate space means the space that consists of more than one point, a Peano continuum is a locally connected continuum, and a continuum is a nonempty connected compact metric space.

\newtheorem{thm}{Theorem}\label{cont}
\begin{thm}
If $X$ is a nondegenerate Peano continuum, for any $\varepsilon>0$, there exist  finitely many infinite Peano subcontinua $X_1,\ldots,X_n$ of $X$ covering $X$ such that dia $X_i~<\varepsilon,~i=1,\ldots,n$. Then, for each $i$, for any positive integer $m^i$, for any $m^i$ points $x^i_1,\ldots,x^i_{m^i}\in X_i$ and $y^i_1,\ldots,y^i_{m^i}\in X$, there exists a continuous onto map $f_{X_i}:X_i\to X$ such that $f_{X_i}(x^i_1)=y^i_1,\ldots,f_{X_i}(x^i_{m^i})=y^i_{m^i}$, and the space $X$, the subsets of $X$, $X_1,\ldots,X_n$, and the maps $f_{X_1},\ldots,f_{X_n}$ satisfy the property {\rm (P)}.
\end{thm}

The nondegenerate Peano continuum such as $n$-cells, $n$-spheres, dendrites, spaces homeomorphic to a torus, spaces homeomorphic to a solid torus, and so on \cite{nad} is ordinarily observed in the real world, phase spaces, and so on, which is both the representation of diverse actual matters and the representation for the realization of various natural phenomena \cite{WEP,Tom}. Then, the relation between the nondegenerate Peano continuum and properties such as hierarchy, self-similarity, and so on was pointed out \cite{O2011}.

In addition, the nondegenerate Peano continuum guarantees infinitely many varieties of the behavior with the property (P) owing to the arbitrariness of the positive number $\varepsilon$ of Theorem \ref{cont}. The above fundamental chaotic maps are simple examples of maps guaranteed from this theorem. 

In this study, let us make a new exploration of sufficient conditions for the property (P).

\section{Sufficient conditions for the property (P)}

Let us start by recalling the following proposition \cite{O2010}, where a topological space $X$ is said to be countably compact provided that every countable open cover of $X$ has a finite subcover.

\newtheorem{pro}{Proposition}
\begin{pro}
If $X$ is a countably compact space, $\{X_\lambda,~\lambda\in\Lambda\}$ is a family of nonempty closed subsets of $X$, and each $f_{X_\lambda}$ is a continuous map from $X_\lambda$ onto $X$, they satisfy the property {\rm (P)}.
\end{pro}
From this proposition, the existence of the behavior with the property (P) is guaranteed by the existence of such a space $X$ and families $\{X_\lambda,~\lambda\in\Lambda\}$ and $\{f_{X_\lambda},~\lambda\in\Lambda\}$.

However, as the existence of the continuous surjections seems to be artificial than the other conditions \cite{O2010} or more essential for the property (P), let us prepare the following proposition which is verified in Appendix for the purpose of the guarantee of this existence, where a topological space $Y$ is perfect provided that  $Y$ contains no isolated points, a topological space $Y$ is zero-dimensional provided that there is a base for the topology of $Y$ such that each element of the base is a closed and open subset of $Y$, and a topological space $Y$ is a $T_1$-space provided that for each $y\in Y$, $\{y\}$ is a closed subset of $Y$.

\begin{pro}
Let $A$ be a zero-dimensional perfect 
compact $T_1$-space. For any compact metric space $X$, there exists a continuous map from $A$ onto $X$.
\end{pro}

From this proposition, if $X$ is a compact metric space and each $X_\lambda$ is a nonempty zero-dimensional perfect closed subset of $X$, there exist continuous surjections $f_{X_\lambda}:X_\lambda\to X,~\lambda\in\Lambda$, and thus they satisfy the property (P) from Proposition 1. Namely, the existence of such a space $X$ and a family $\{X_\lambda,~\lambda\in\Lambda\}$ guarantees the existence of the behavior with the property (P).

However, this time, as the conditions of the family $\{X_\lambda,~\lambda\in\Lambda\}$ of Proposition 2 seems to be artificial or essential, let us further prepare the following proposition for the guarantee of the existence, where a $T_0$-space is a topological space $Y$ such that for any points $y_1$ and $y_2$ in $Y$ with $y_1\ne y_2$, there exists an open subset $U$ of $Y$ such that $y_i\in U$ and $y_j\notin U$ for some choice of $i$ and $j$.

\begin{pro}
Let $X$ be a zero-dimensional perfect $T_0$-space. For any positive integer $n$, there exist nonempty closed and open subsets $X_1,\ldots,X_n$ of $X$ such that $X_i\cap X_{i'}=\emptyset,~i\ne i',~X_1\cup \cdots\cup X_n=X$; that is, $\{X_1,\ldots,X_n\}$ is a partition of $X$ each element of which is a closed and open subset of $X$.
\end{pro}

{\noindent\bf Proof}~~Let us use the mathematical induction. Let $\{X_1,\ldots,X_{n-1}\}$ be a partition of $X$ each element of which is a closed and open subset of $X$. From the perfectness of $X$, $X_{n-1}$ contains two different points $x$ and $y$, because if $X_{n-1}$ is a singleton $\{x\}$, $x$ is an isolated point in $X$. 

Without loss of generality, there exists an open subset $u$ of $X$ such that $x\in u$ and $y\notin u$, because $X$ is a $T_0$-space. Since $X$ is zero-dimensional, there exists a  closed and open subset $v$ of $X$ such that $x\in v\subset u\cap X_{n-1}$. Since $X_{n-1}-v$ contains $y$ and is a closed and open subset of $X$, the desired partition $\{X_1,\ldots,X_{n-2},X_{n-1}-v,v\}$ is obtained. $\Box$\\

Here, any nonempty open subset $A$ of a perfect space $Y$ is perfect, because if $a$ is an isolated point in $A$, $a$ is also an isolated point in $Y$. Then, any nonempty subset $A$ of a zero-dimensional space $Y$ is zero-dimensional. This is because if $\{U_\lambda,~\lambda\in\Lambda\}$ is a base for the topology of $Y$ each element of which is a closed and open subset of $Y$, $\{U_\lambda\cap A,~\lambda\in\Lambda\}$ is a base for the topology of the subspace $A$ of $Y$ each element of which is a closed and open subset of $A$.


Accordingly, if $X$ is a zero-dimensional perfect compact metric space, for any positive integer $n$, there exists a partition $\{X_1,\ldots,X_n\}$ of $X$ each element of which is a zero-dimensional perfect closed subsets of $X$. From Proposition 2, there exist continuous surjections $f_{X_i}:X_i\to X,~i=1,\ldots,n$, and the existence of the behavior with the property (P) is guaranteed from Proposition 1. 

Furthermore, lecalling that a space is a Cantor set if and only if it is a zero-dimensional perfect compact space \cite[p.65]{IN}, where any space that is homeomorphic to the Cantor middle-third set is called a Cantor set, we obtain the following theorem.


\begin{thm}
If $X$ is a Cantor set, for any positive integer $n$, there exist a partition $\{X_1,\ldots,X_n\}$ of $X$ each element of which is a closed and open subset of $X$ and continuous surjections $f_{X_i}:X_i\to X,~i=1,\ldots,n$. Then, they satisfy the property {\rm (P)}.
\end{thm}

In addition, the Cantor set guarantees infinitely many varieties of the behavior with the property (P), as well as the nondegenerate Peano continuum, owing to the arbitrariness of the positive integer $n$ of this theorem. In particular, the map $f:S\to S,~x\mapsto 3x~(x\in [0,1/3]\cap S),~ 3x-2~(x\in [2/3,1]\cap S)$ is a simple example of maps guaranteed from this theorem where $S$ denotes the Cantor middle-third set, and the property (P) can play an important role in the guarantee of nonperiodicity, sensitive dependence on initial conditions, and so on for this map.

\section{Conclusions}

It is exhibited that a Cantor set homeomorphic to the Cantor set which is a fundamental fractal emerges as a result of the exploration of sufficient conditions for the property (P) generalized from a property specific to fundamental chaotic maps, and the Cantor set even guarantees infinitely many varieties of the behavior with the property (P).

It is interesting that the contrast spaces, the Cantor set and the nondegenerate Peano continuum, emerge by the exploration of the same property (P); the Cantor set is totally disconnected \cite[Theorem 7.14]{nad} while the nondegenerate Peano continuum is connected and locally connected by definition.


\section*{Acknowledgments}

The authors are grateful to Professor Akihiko Kitada of Waseda University and Professor Emeritus Yoshisuke Ueda of Kyoto University for useful discussions. This study was supported by the Japan Science and Technology Agency.


\appendix

\section{Proof of Proposition 2}

Since $X$ is a compact metric space, there exist finitely many open balls $S_1,\ldots,S_n$ such that $S_1\cup\cdots\cup S_n=X$ and $dia\,S_i$~(the diameter of $S_i$)~$<1/2$ for each $i$. Letting $X_i$ be the closure of $S_i$ in $X$ for each $i$, we obtain nonempty closed subsets $X_1,\ldots,X_n$ of $X$ such that $X_1\cup\cdots\cup X_n=X$ and $dia\,X_i<1$ for each $i$.

From Proposition 3, there exist a partition $\{A_1,\ldots,A_n\}$ of $A$ each element of which is a closed and open subset of $A$. Then, the map
\begin{eqnarray}
g_1:A\to \Im,~a\mapsto X_i~(a\in A_i)
\end{eqnarray}
is obtained, where $\Im$ denotes the family of all closed subsets of $X$.

Since each subspace $X_i$ of $X$ is a compact metric space, there exist finitely many open balls $S_{i1},\ldots,S_{in_i}$ of $X_i$ such that $S_{i1}\cup\cdots\cup S_{in_i}=X_i$ and $dia\,S_{ij}<1/4$ for each $j$. Letting $X_{ij}$ be the closure of $S_{ij}$ in $X_i$ for each $j$, we obtain nonempty closed subsets  $X_{i1},\ldots,X_{in_i}$ of $X$ such that $X_{i1}\cup\cdots\cup X_{in_i}=X_i$ and $dia\,X_{ij}<1/2$.

From Proposition 3, for each $i$, there exist a partition $\{A_{i1},\ldots,A_{in_i}\}$ of $A_i$ each element of which is a closed and open subset of $A$. Then, the map
\begin{eqnarray}
g_2:A\to \Im,~a\mapsto X_{ij}~(a\in A_{ij})
\end{eqnarray}
is obtained.

Since this procedure can be repeated, the family $\{g_k:A\to \Im,~k\in\mathbb{N}\}$ of the surjections such that for each $k$, 
\begin{eqnarray}
dia\,g_k(a)<\frac{1}{k},~^\forall a\in A,
\end{eqnarray}
is obtained. Since $\bigcap_{k=1}^\infty g_k(a)$ is a singleton for each $a\in A$, the map 
\begin{eqnarray}
G:A\to\{\{x\};~x\in X\},~a\mapsto\bigcap_{k=1}^\infty g_k(a) 
\end{eqnarray}
is obtained. 

Next, let us verify that the map
\begin{eqnarray}
p\circ G:A\to X 
\end{eqnarray}
is continuous, where 
\begin{eqnarray}
p:\{\{x\};~x\in X\}\to X,~\{x\}\mapsto x.
\end{eqnarray}
For any $a\in A$ and for any open subset $U$ of $X$ such that $p\circ G(a)\in U$, since $\bigcap_{k=1}^\infty g_k(a)\subset U$, there exists $k_0$ such that $g_{k_0}(a)\subset U$ from Lemma 2. By the definition of $g_{k_0}$, there exists an open (and closed) subset $u$ of $A$ such that $a\in u$ and $g_{k_0}(a')\subset U$ for any $a'\in u$. Since $G(a')\subset g_{k_0}(a')\subset U$ for any $a'\in u$, $p\circ G(u)\subset U$, and thus $p\circ G$ is continuous.

Lastly, let us verify that $p\circ G$ is a surjection.
For any $x\in X$, there exists $a_k\in A$ such that $x\in g_k(a_k)$ for each $k$, by the definition of $g_k$. Accordingly, the sequence $\{a_k\}_{k=1}^{\infty}$ of points in $A$ such that $x\in g_k(a_k)$ for each $k$, is obtained. 

Let us consider a case such that $\{a_k\}_{k=1}^{\infty}$ is a finite set. Since there exists a number $K$ such that $a_K=a_k$ for any $k\ge K$, $G(a_K)=\bigcap_{k=1}^\infty g_k(a_K)=\{x\}$. Namely, $p\circ G(a_K)=x$. 

Let us consider a case such that  $\{a_k\}_{k=1}^{\infty}$ is an infinite set. Since $A$ is compact, there exists an accumulation point $a$ of $\{a_k\}_{k=1}^{\infty}$. If $G(a)\ne\{x\}$, $G(a)\subset X-\{x\}$, and thus there exists $k_0$ such that $g_{k_0}(a)\subset X-\{x\}$ from Lemma 2. Accordingly, there exists an open (and closed) subset $u$ of $A$ such that $a\in u$ and $g_{k_0}(a')\subset X-\{x\}$ for any $a'\in u$. From Lemma 1, there exists an open subset $u'$ of $A$ such that $a\in u'\subset u$ and $u'\cap (\{a_k\}_{k=1}^{\infty}-\{a_1,\ldots,a_{k_0}\})\ne\emptyset$. Therefore, there exists $a_{k'}\in u'$ $(k'> k_0)$, and thus $x\in g_{k'}(a_{k'})\subset g_{k_0}(a_{k'})\subset X-\{x\}$. This is a contradiction, and thus $G(a)=\{x\}$. Namely, $p\circ G$ is a surjection. $\Box$

\newtheorem{lem}{Lemma}
\begin{lem}
Let $Y$ be a $T_1$-space, $A$ be an infinite subset of $Y$, and $a$ be an accumulation point of $A$. For any finite subset $B$ of $A$ and for any open subset $U$ of $Y$ such that $a\in U$, there exists an open subset $u$ of $Y$ such that $a\in u\subset U$ and $u\cap (A-B)\ne\emptyset$.
\end{lem}

{\noindent\bf Proof}~~Since $Y$ is a $T_1$-space, there exists an open subset $U'$ of $Y$ such that $a\in U'$ and $U'\cap B=\emptyset$. Since $a$ is an accumulation point of $A$, $(U\cap U')\cap A\ne\emptyset$, and thus $(U\cap U')\cap (A-B)\ne\emptyset$. $\Box$\\

\begin{lem}
Let $Y$ be a compact $T_1$-space and $\{A_i\}$ be the sequence of nonempty closed subsets of $Y$ such that $A_i\supset A_{i+1}$ for each $i$. For any open subset $U$ of $Y$ such that $\bigcap_{i=1}^{\infty} A_i\subset U$, there exists $i_0$ such that $A_{i_0}\subset U$.
\end{lem}

{\noindent\bf Proof}~~If there exists an open subset $U$ of $Y$such that $\bigcap_{i=1}^{\infty} A_i\subset U$ and $A_{i}\not\subset U$ for any $i$, there exists a sequence $\{a_i\}_{i=1}^{\infty}$ of points in $A$ such that $a_i\in A_i\cap (X-U)$ for each $i$. 

Let us consider a case such that $\{a_i\}_{i=1}^{\infty}$ is a finite set. Since there exists a number $I$ such that $a_I=a_i$ for any $i\ge I$, $a_I\in\bigcap_i A_i\subset U$. This is a contradiction. 

Let us consider a case such that $\{a_i\}$ is an infinite set. Since $X$ is compact, there exists an accumulation point $a$ of $\{a_i\}_{i=1}^{\infty}$. If $a\in U$, there exists $i'$ such that $a_{i'}\in U$. This is a contradiction. Thus,
\begin{eqnarray}
a\in X-U.\label{Uc} 
\end{eqnarray}

Let us verify that $a$ is contained in the closure of $A_{i'}$ for each $i'$.  For any open subset $u$ of $Y$ such that $a\in u$, there exists an open subset $u'$ of $Y$ such that $a\in u'\subset u$ and $u'\cap(\{a_i\}_{i=1}^{\infty}-\{a_1,\ldots,a_{i'}\})\ne\emptyset$, from Lemma 1. Accordingly, $u\cap A_{i'}\supset u'\cap(\{a_i\}_{i=1}^{\infty}-\{a_1,\ldots,a_{i'}\})\ne\emptyset$, and thus $a$ is contained in the closure of $A_{i'}$. 

Since each $A_{i'}$ is closed, $a\in\bigcap_{i'} A_{i'}\subset U$. This is in contradiction with the relation (\ref{Uc}). $\Box$



\bibliographystyle{model1a-num-names}
\bibliography{<your-bib-database>}







\end{document}